\documentclass[sigconf,9pt]{acmart}

\AtBeginDocument{%
  }
    
\setcopyright{none} 
\renewcommand\footnotetextcopyrightpermission[1]{}
    
\usepackage{subcaption}
\usepackage{xspace}
\usepackage{paralist}

\newcommand{\ambiSQL}{\textsc{AmbiSQL}\xspace}
\newcommand{\circled}[1]{\textcircled{\small #1}}

\newcommand\blfootnote[1]{%
  \begingroup
  \renewcommand\thefootnote{}\footnote{#1}%
  \addtocounter{footnote}{-1}%
  \endgroup
}

\acmConference[SIGMOD-Companion '26]{}{May 31--June 5,
  2026}{Bengaluru, India}

\begin{document}

\title{\ambiSQL: Interactive Ambiguity Detection and Resolution for Text-to-SQL}

\author{
Zhongjun Ding, 
Yin Lin$^\dag$,
Tianjing Zeng,
Rong Zhu$^\dag$,
Bolin Ding,
Jingren Zhou
}
\affiliation{%
Alibaba Group\\
\vspace{1mm}
\texttt{\small \{dingzhongjun.dzj, yin.lin, zengtianjing.ztj, red.zr, bolin.ding, jingren.zhou\}@alibaba-inc.com}
\country{}
}

\renewcommand{\shortauthors}{Zhongjun Ding, Yin Lin,
Tianjing Zeng, Rong Zhu, Bolin Ding, Jingren Zhou}

\begin{abstract}
Text-to-SQL systems translate natural language questions into SQL queries, providing substantial value for non-expert users. While large language models (LLMs) show promising results for this task, they remain error-prone. \emph{Query ambiguity} has been recognized as a major obstacle in LLM-based Text-to-SQL systems, leading to misinterpretation of user intent and inaccurate SQL generation. To this end, we present \ambiSQL, an interactive system that automatically detects query ambiguities and guides users through intuitive multiple-choice questions to clarify their intent. It introduces a fine-grained ambiguity taxonomy for identifying ambiguities arising from both database elements and LLM reasoning, and subsequently incorporates user feedback to rewrite ambiguous questions. 

In this demonstration, \ambiSQL is integrated with XiYan-SQL, our commercial Text-to-SQL backend. We provide 40 ambiguous queries collected from two real-world benchmarks that SIGMOD'26 attendees can use to explore how disambiguation improves SQL generation quality. Participants can also apply the system to their own databases and natural language questions. The codebase and demo video are available at: \url{https://github.com/JustinzjDing/AmbiSQL} and \url{https://www.youtube.com/watch?v=rbB-0ZKwYkk}.
\end{abstract}

\maketitle
\blfootnote{$^\dag$ Corresponding authors.}
\vspace{-2em}
\section{Introduction}
\label{sec: intro}
Natural language interfaces for databases have been extensively studied in the database community. Recent advances \cite{xiyan-sql} in applying large language models (LLMs) for Text-to-SQL translation have led to remarkable improvements, as evidenced by their performance on widely used benchmarks such as Spider~\cite{DBLP:conf/emnlp/YuZYYWLMLYRZR18} and BIRD~\cite{DBLP:conf/nips/LiHQYLLWQGHZ0LC23}.

Despite these promising results, \textit{query ambiguity} remains a key error source in LLM-based Text-to-SQL systems, causing misalignment between the user's actual intent and generated SQL queries \cite{cidr26-broken-text2sql, sphinteract, DBLP:conf/cidr/FloratouPZDHTCA24}.
Notably, such ambiguity is pervasive and often unavoidable in real-world settings, since no end-to-end system can ensure all user queries are perfectly precise and unambiguous.
Given an ambiguous query, these systems might generate syntactically valid but semantically inaccurate SQL queries that do not meet the user's expectations. This creates significant challenges for Text-to-SQL system development (e.g., difficulty in linking natural language questions to the correct database elements), evaluation (e.g., queries may have multiple valid interpretations while benchmarks provide only a single ground truth answer), and real-world deployment (e.g., systems struggle to accurately parse and understand user intentions without additional context).

Query ambiguities manifest in various forms, each presenting unique challenges for accurate SQL generation.
Consider a query from BIRD \cite{DBLP:conf/nips/LiHQYLLWQGHZ0LC23}: \textit{``What is the average number of test takers from Fresno schools that opened between 01/01/1980 and 12/31/1980?''} The geographical reference ``Fresno'' can be mapped to the ``City'' or ``County'' column in the database. Users need to explicitly specify ``schools in Fresno County'' to resolve this ambiguity.
Queries requiring complex reasoning and external knowledge present additional ambiguities. A question from the TAG benchmark \cite{DBLP:journals/corr/abs-2408-14717} \textit{``How many drivers born after the end of the Vietnam War have been ranked 2?''} contains two unclear references: the temporal constraint ``end of the Vietnam War'' may be interpreted at different levels of granularity (year vs. exact date), and ``ranked 2'' can be mapped to the ``position'' or ``rank'' columns across different tables.

To enable more reliable SQL generation from natural language queries, we perform ambiguity identification and query refinement through user clarifications. Our approach centers on a comprehensive taxonomy of fine-grained ambiguity types to detect and resolve issues that affect \textit{database element mapping} and \textit{LLM reasoning}. Guided by this taxonomy, we use in-context learning to automatically detect ambiguous phrases, classify them by type, and generate targeted clarification questions (CQs) with intuitive multiple-choice options. User selections then guide automatic query rewriting to resolve the identified ambiguities.

We demonstrate our approach through \ambiSQL, a user-friendly demo system integrated with our commercial Text-to-SQL system XiYan-SQL. \ambiSQL provides an interactive UI featuring 40 ambiguous queries curated from the BIRD~\cite{DBLP:conf/nips/LiHQYLLWQGHZ0LC23} and TAG~\cite{DBLP:journals/corr/abs-2408-14717} benchmarks. These examples demonstrate how \ambiSQL helps users, even without SQL or database knowledge, to interact with the system through simple multiple-choice selections to generate more accurate SQL queries. Users can also apply the system to analyze their own datasets and natural language questions.

\textit{Related Work.}
Existing disambiguation approaches follow two main strategies: engaging users for clarification after ambiguity identification~\cite{DBLP:journals/pacmmod/ChenCKY25,sphinteract,clear}, or leveraging LLMs to automatically generate and select among candidate interpretations~\cite{xiyan-sql, saparina2025disambiguateparselatergenerating, gong2025sqlen}. Despite their effectiveness, these approaches still face several limitations. (1) Fully automated approaches~\cite{saparina2025disambiguateparselatergenerating,xiyan-sql,gong2025sqlen} may diverge from the user’s true intent because ambiguities are resolved without explicit user feedback. (2) More broadly, both human-in-the-loop and automated methods~\cite{DBLP:journals/pacmmod/ChenCKY25,saparina2025disambiguateparselatergenerating,sphinteract,clear} predominantly target schema-linking errors and data-related ambiguities, leaving ambiguities that primarily affect LLM reasoning largely unexplored (e.g., temporal interpretations of phrases such as ``the end of the Vietnam War''). (3) Finally, in the absence of a comprehensive ambiguity taxonomy, many systems generate an excessive number of candidate interpretations~\cite{saparina2025disambiguateparselatergenerating,xiyan-sql,gong2025sqlen}, leading to substantial computational overhead. Approaches based on pre-trained ambiguity detectors~\cite{DBLP:journals/pacmmod/ChenCKY25,saparina2025disambiguateparselatergenerating} often require costly annotations and tend to generalize poorly across domains.

\section{System Architecture}
\subsection{Ambiguity Taxonomy} 
\label{sec:taxonomy}

We classified the ambiguity types observed during Text-to-SQL evaluation. These ambiguities cause misalignments between natural language questions and database elements, as well as LLM reasoning errors, particularly when external knowledge evidence is incorporated during SQL generation~\cite{DBLP:conf/nips/LiHQYLLWQGHZ0LC23} or LLM-based knowledge retrieval is required during query execution~\cite{DBLP:journals/corr/abs-2408-14717}.

The first dimension is \textbf{\textit{Database-sourced ambiguity}}: Ambiguity arising from unclear or underspecified references to database schema or content in natural language questions, leading to incorrect or incomplete data retrieval. This includes:

\vspace{-0.2em}
\begin{itemize}
\item \textbf{\textsc{AmbiSchema}\xspace}: The question lacks sufficient context to determine which table or column to use for operations like filtering, ranking, or aggregation, resulting in multiple plausible interpretations (e.g., ``oldest user'' could refer to age or registration date). \textsc{AmbiSchema}\xspace leads to incorrect database source mapping in the logical plan.

\item \textbf{\textsc{AmbiValue}\xspace}: The question refers to a value that does not correspond to actual values stored in the database, making it unclear how to form the correct literals in the WHERE clause. This can cause relevant records to be omitted or produce inaccurate results (e.g., querying for ``New York City'' when the database stores ``NYC,'' or asking for posts about ``COVID-19'' when the database contains ``coronavirus'').

\item \textbf{\textsc{AmbiIntent}\xspace}: Key terms clarifying the intended SQL operation are absent, creating ambiguity about the desired operation (e.g., ``Show me users by registration date'' could mean ORDER BY for sorting, GROUP BY for grouping, or WHERE for filtering).

\end{itemize}
\vspace{-0.2em}

The second dimension is \textbf{\textit{LLM-sourced ambiguity}}: Ambiguity arising from unclear or underspecified requirements for LLM reasoning, causing difficulties in correctly retrieving or applying information beyond the database content. This includes:

\vspace{-0.2em}
\begin{itemize}
\item \textbf{\textsc{AmbiSource}\xspace}: 
 The question does not specify whether relevant data should be retrieved from the database or inferred via LLM reasoning (e.g., querying for ``players taller than Michael Jordan'', where it is unclear if Jordan's height should come from the database or an external data source).

\item \textbf{\textsc{AmbiContext}\xspace}: The question lacks adequate context to guide LLM reasoning effectively (e.g., requesting ``exchange rate'' without specifying the target currencies or date).

\item \textbf{\textsc{AmbiFallacy}\xspace}: Knowledge assumptions embedded within the question contradict real-world facts or database contents (e.g., querying participants in events that never occurred).

\item \textbf{\textsc{AmbiRef}\xspace}: Spatial or temporal constraints are underspecified, resulting in multiple plausible interpretations at different granularities (e.g., ``after the 2018 World Cup'' could mean exactly after the final match or after the tournament year).
\end{itemize}
\vspace{-0.2em}

\subsection{Problem Formulation}

We model a Text-to-SQL system as a function $f: ({\sf q}, \mathcal{D}) \mapsto \mathcal{Q}$ that maps a natural language question ${\sf q}$ and a database schema $\mathcal{D}$ to a SQL query $\mathcal{Q}$.
We define a set of ambiguity types $\mathcal{A} = \{a_1, a_2, \ldots, a_n\}$ based on our taxonomy, where each $a_i$ represents a distinct ambiguity category that may introduce errors in Text-to-SQL.
In \ambiSQL, we address the following two problems:

\textbf{Ambiguity detection:} 
Given a natural language question ${\sf q}$, a database schema $\mathcal{D}$, a Text-to-SQL system $f$, and the predefined ambiguity types $\mathcal{A}$, identify (1) the set of ambiguous phrases $P = \{p_1, p_2, \dots, p_k\}$ within ${\sf q}$, where each $p_i$ is a substring of ${\sf q}$, and (2) the corresponding ambiguity type $a_i \in \mathcal{A}$ for each phrase $p_i$, such that these ambiguous phrases cause $f({\sf q}, \mathcal{D})$ to generate a SQL query $\mathcal{Q} \neq \mathcal{Q}^*$, where $\mathcal{Q}^*$ denotes the user's query intent. 

\textbf{Ambiguity resolution:} Given a natural language question ${\sf q}$ with identified ambiguous phrases $P = \{p_1, p_2, \dots, p_k\}$ and their ambiguity types $\{a_1, a_2, \dots, a_k\}$, generate a clarified question ${\sf q}'$ such that $f({\sf q}', \mathcal{D}) = \mathcal{Q}^*$.

\begin{figure}
    \centering
    \includegraphics[width=1.0\linewidth]{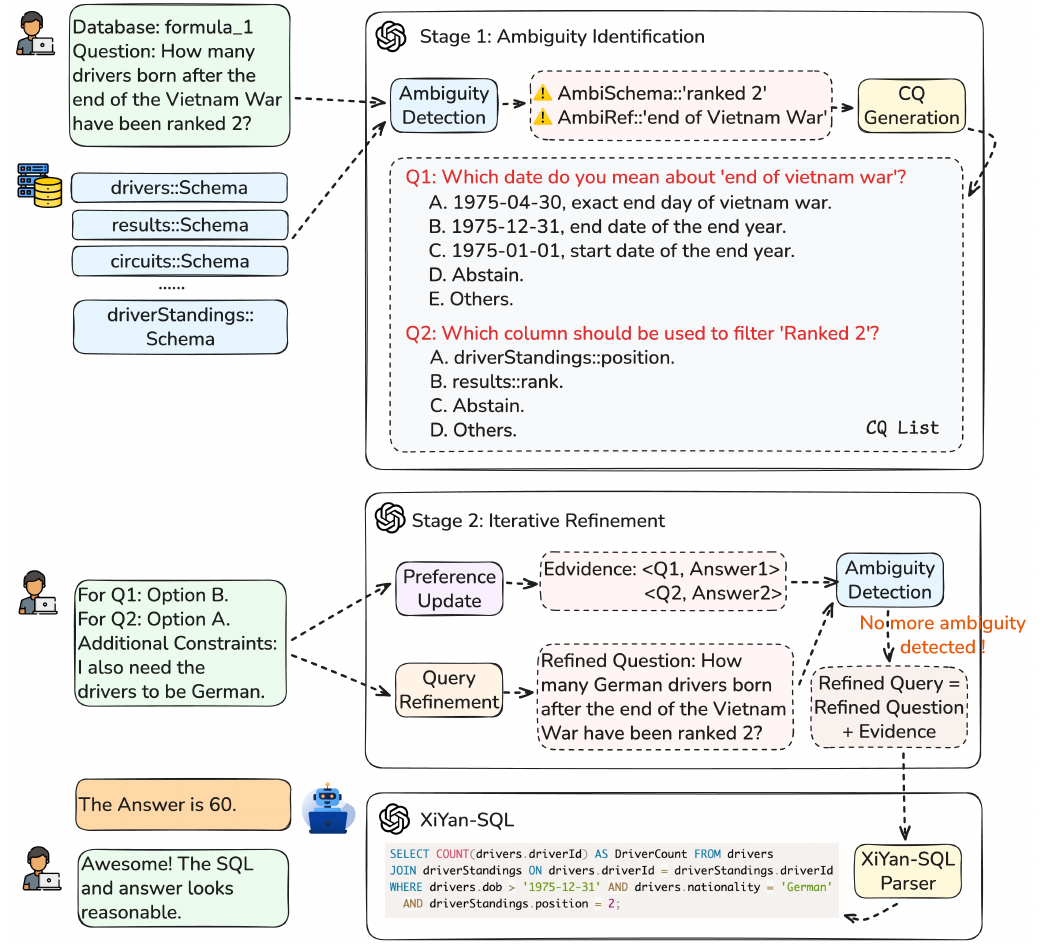}
    \vspace{-5mm}
    \caption{\textnormal{An Overview of Ambiguity Detection and Resolution in \ambiSQL.} 
    }
    \label{fig:system overview}
    \vspace{-5mm}
\end{figure}

\vspace{-0.5em}
\subsection{System Overview}
We present an overview of the \ambiSQL system in Figure~\ref{fig:system overview}. The system employs a two-stage pipeline: \textit{Ambiguity Identification} and \textit{Iterative Refinement}. The first stage aims to identify ambiguities in the original user query, while the second stage focuses on iteratively refining the query via user clarifications. 

\textbf{Stage 1: Ambiguity Identification.}
Given a natural language query and the corresponding database schema, the system first employs the ambiguity detection module to identify potentially ambiguous phrases (e.g., ``ranked 2'' and ``end of the Vietnam War'' in our running example), and then classifies them according to our ambiguity taxonomy. These identified ambiguities are subsequently passed to the clarification question generation module, which formulates each ambiguous aspect into a targeted clarification question (CQ), often supplemented by relevant descriptions or database schema snippets (cf. grey background panel in Stage 1). The generated clarification questions, along with their accompanying descriptions, are then presented to the user for response.

\textbf{Stage 2: Iterative Refinement.}
After the user provides answers to the clarification questions, both the questions and their answers are processed by the preference update module and used as inputs for Stage 2.
Additionally, \ambiSQL allows users to specify \textit{additional constraints} (e.g., ``drivers need to be German'') as extra conditions to refine the original query and clarify their intent. The query refinement module integrates the original query with any additional constraints to produce an updated query. 
This refined query, together with the user clarifications, is then re-examined by the ambiguity detection module to ensure no new ambiguities were introduced during the clarification process. Once all ambiguities are resolved, the refined query and user clarifications are submitted to the underlying Text-to-SQL system, which generates a SQL query reflecting the user's intent.

Section~\ref{sec:design detail} provides detailed descriptions of all core system modules: the ambiguity detection module (used in both stages), the clarification question generation module (Stage 1), and the preference update and query refinement modules (Stage 2).

\vspace{-0.5em}
\subsection{Technical Details}
\label{sec:design detail}
\textbf{Ambiguity Detection.} Recent work~\cite{DBLP:conf/cidr/FloratouPZDHTCA24} reveals that advanced LLMs achieve agreement rates with human annotators comparable to inter-annotator agreement on ambiguity detection tasks. Building on this observation, \ambiSQL employs pre-trained LLMs with in-context learning to identify ambiguities in user questions. 

The key challenge is instructing the LLM to detect ambiguous phrases and accurately categorize them according to our taxonomy. We address this through carefully designed prompts that include: (1) explicit ambiguity definitions, (2) our concise taxonomy from Section~\ref{sec:taxonomy}, and (3) representative examples for each ambiguity type. For instance, for the question ``Which city is the largest one?'', we provide an example explanation: \textit{``This exhibits \textsc{AmbiSchema}\xspace—the user does not specify whether ``largest'' refers to area or population.''} With such guidance, the LLM can generalize to recognize similar ambiguities in new questions.

When ambiguity dependencies exist in the question—for example, when we need to resolve \textsc{AmbiIntent}\xspace to determine the correct operator before detecting \textsc{AmbiSchema}\xspace and \textsc{AmbiValue}\xspace related to database element usage—the iterative refinement in Stage 2 ensures that all dependent ambiguities are detected and resolved.

\textbf{Clarification Question (CQ) Generation.}
After detecting ambiguous phrases and their types, \ambiSQL generates targeted multiple-choice questions to help users clarify their intent. For each ambiguity, the system creates a descriptive question with options informed by relevant database content or contextual knowledge. 
For database-sourced ambiguities, we present relevant database snippets and ask users to select the intended schema element, value, or operation. For LLM-sourced ambiguities, we ask whether the required information should be retrieved from the database or inferred via LLM reasoning, and request additional context when necessary.

Each clarification question includes two default options: \textit{Abstain} and \textit{Others}. The \textit{Abstain} option allows users to skip clarification when the generated question is not relevant to their query, and the system uses this feedback to improve future question generation. The \textit{Others} option allows users to enter their clarifications freely to handle cases where the provided multiple-choice options do not include the user's intended answer.

\textbf{Preference Update.}
\ambiSQL stores user clarifications in a tree-like structure organized by our ambiguity taxonomy. Each ambiguity type has a corresponding node that stores the user's clarification responses. For example, when a user clarifies \textsc{AmbiSchema}\xspace by selecting a specific column, this preference is recorded under the corresponding taxonomy node. When a user's new clarification conflicts with a previously stored preference, the system automatically updates the stored preference to reflect the latest intent. This ensures that the system always uses the most current user preferences when resolving ambiguities.

\textbf{Query Refinement.}
This module handles users' \textit{additional constraints} by synthesizing them into the original query. When conflicts arise between the original and additional constraints, the system prioritizes the additional constraints to reflect the user's most recent intent.
Since additional constraints may themselves be ambiguous, we handle them separately from the preference update module and also perform ambiguity detection on the rewritten query.

\begin{figure*}
    \centering
    \includegraphics[width=1.0\linewidth]{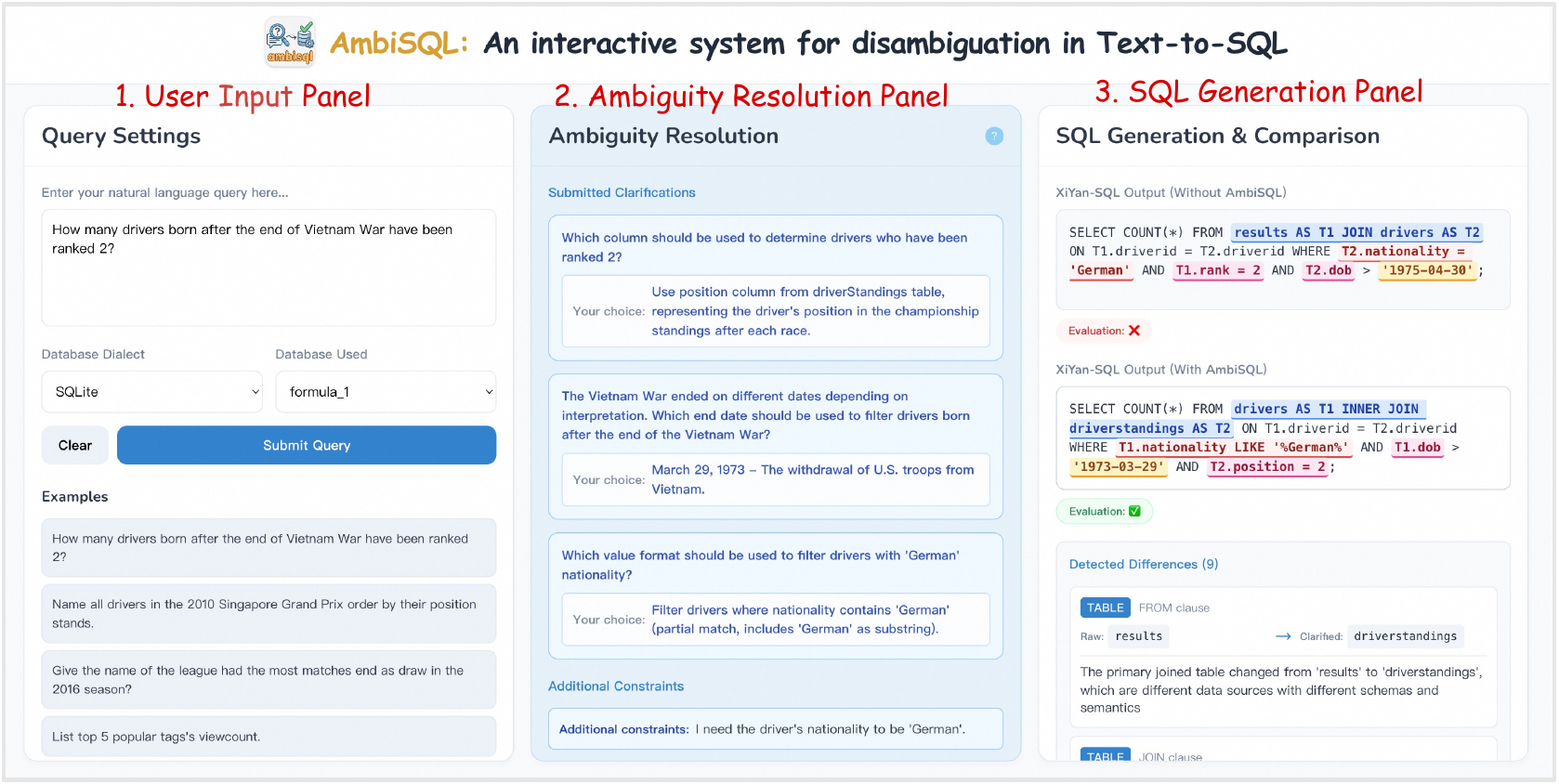}
    \vspace{-2mm}
    \caption{\textnormal{Demonstration of \ambiSQL.} 
    }
    \label{fig:demo ui}
    \vspace{-3mm}
\end{figure*}

\vspace{-0.5em}
\section{Demonstration}
\label{sec:demonstration}

We integrate \ambiSQL with our commercial Text-to-SQL backend, XiYan-SQL~\cite{xiyan-sql}, to deliver an end-to-end interactive demonstration. We provide a playground dataset with 40 carefully curated ambiguous queries collected from BIRD~\cite{DBLP:conf/nips/LiHQYLLWQGHZ0LC23} and TAG~\cite{DBLP:journals/corr/abs-2408-14717}, enabling attendees to quickly explore ambiguity cases. This dataset covers all ambiguity types defined in Section~\ref{sec:taxonomy}, with each query accompanied by human-annotated ambiguity categories and actual user intent. In addition to the built-in playground, \ambiSQL also supports user-customized queries and databases, allowing attendees to evaluate the system with their own scenarios. Participants can explore these features through a user interface with three functional panels (\circled{1}–\circled{3}), as shown in Figure~\ref{fig:demo ui}.

\textbf{User Input Panel \circled{1}.} 
To initiate the \ambiSQL process, users can either input a custom natural language query or select one example from the playground dataset. This panel accepts three primary parameters: the natural language query, the target SQL dialect, and the specific database to be queried. Selecting an example automatically populates these fields, streamlining the exploration.

\begin{figure}[t]
  \centering
  \begin{subfigure}{0.49\linewidth}
    \centering
    \includegraphics[width=\linewidth]{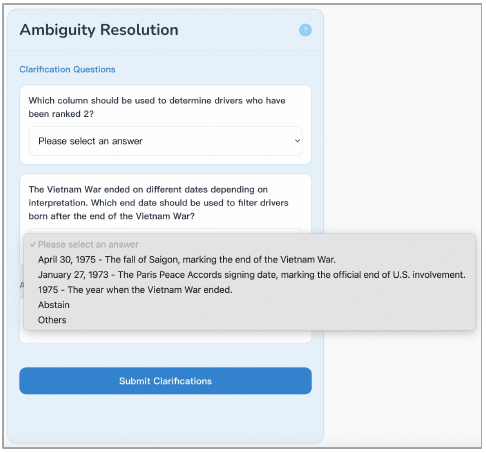}
    \caption{\textnormal{Ambiguity resolution interface via multiple-choice questions.}}
    \label{fig:choice card}
  \end{subfigure}\hfill
  \begin{subfigure}{0.49\linewidth}
    \centering
    \includegraphics[width=\linewidth]{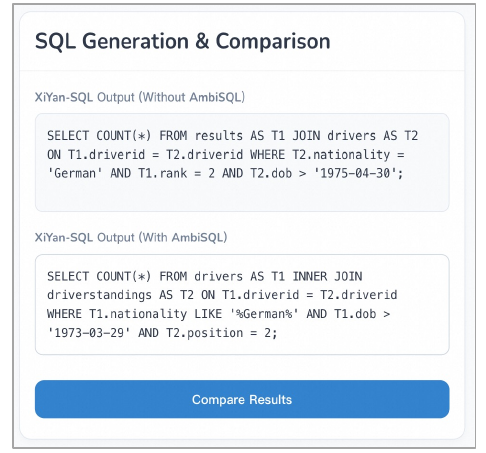}
    \caption{\textnormal{Side-by-side comparison of generated SQL with and without \ambiSQL.}}
    \label{fig:sql comparison}
  \end{subfigure}
  \vspace{-0.6em}

  \caption{\textnormal{Detailed User Interface of \ambiSQL.}}
  \label{fig:two}
  \vspace{-0.8em}
\end{figure}

\textbf{Ambiguity Resolution Panel \circled{2}.}
Upon submission, \ambiSQL analyzes the user inputs for ambiguity detection (Stage~1 in Figure~\ref{fig:system overview}). For each identified ambiguity, the system generates a multiple-choice clarification question with candidate options accessible via dropdown menus, as shown in Figure~\ref{fig:choice card}.

As illustrated in Figure~\ref{fig:demo ui}, the phrase ``the end of the Vietnam War'' is disambiguated by presenting plausible concrete historical dates (e.g., April 30, 1975 vs. March 29, 1973). Alongside, the \textsc{AmbiSchema}\xspace ambiguity with respect to the phrase ``ranked 2'' is addressed by providing candidate schema elements, such as the \texttt{``position''} column in the \texttt{``driverStandings''} table with brief descriptive snippets. Additionally, an \textit{Additional Constraints} field enables users to supply auxiliary instructions (e.g., “the driver’s nationality should be German”) to further specify the original query.

After users submit their clarifications, \ambiSQL iteratively refines the query and identifies any remaining ambiguities through multi-round interactions, continuing until no unresolved ambiguity is detected. The resulting clarified, unambiguous rewrite is then forwarded to XiYan-SQL for SQL generation. For traceability, Panel \circled{2} maintains a persistent interaction log that records all user-provided clarifications and additional constraints, enabling users to track how each ambiguity is resolved step by step.

\textbf{SQL Generation Panel \circled{3}.} Once all ambiguities have been resolved through interaction, Panel \circled{3} is activated to generate SQL via XiYan-SQL. As shown in Figure~\ref{fig:sql comparison}, we present the Text-to-SQL outputs with and without \ambiSQL side by side for comparison. In the illustrated example, both outputs correctly incorporate the core intent of the original query as well as the additional constraint: they retain the original filtering conditions and further add a user-specified filter on the driver’s nationality.

The \emph{Compare Results} button validates each generated SQL against the pre-defined query intent and contrasts the two outputs in terms of SQL structure and referenced database entities. This comparison reveals the impact of ambiguity resolution. In particular, the \textit{Detected Differences} sub-panel pinpoints the exact refinements introduced after clarification. For example, \ambiSQL correctly maps the phrase “ranked 2” to the \texttt{position} column and updates the temporal filter from \texttt{1975-04-30} to \texttt{1973-03-29} based on user feedback. By proactively resolving such ambiguities, \ambiSQL enables the downstream model to generate SQL that more faithfully and consistently reflects the user’s intent.

\section{Conclusion} 
We demonstrate how \ambiSQL detects and resolves ambiguity in Text-to-SQL systems via a human-in-the-loop clarification process. Leveraging a systematic, fine-grained ambiguity taxonomy and a two-stage clarification pipeline, \ambiSQL rewrites natural language questions to better align with users’ actual intent. We hope our system can inspire SIGMOD’26 attendees to explore new opportunities for improving both Text-to-SQL systems and the robustness of related benchmarks.

\bibliographystyle{ACM-Reference-Format}
\bibliography{ref}
\end{document}